\documentclass[a4paper]{article}
\usepackage{bm}
\usepackage{amsfonts}
\usepackage[english]{inputenc}
\usepackage{amsmath}
\usepackage{amssymb}
\usepackage{amsthm}
\usepackage{authblk}
\usepackage{color}

\begin{document}

\newtheorem{definition}{Definiton}[section]
\newtheorem{theorem}{Theorem}[section]

\title{Uniqueness theorem for static phantom wormholes in Einstein-Maxwell-dilaton theory}
\author[1]{Boian Lazov\thanks{boian\_lazov@phys.uni-sofia.bg}}
\author[1,2]{Petya Nedkova\thanks{pnedkova@phys.uni-sofia.bg}}
\author[1,3]{Stoytcho Yazadjiev\thanks{yazad@phys.uni-sofia.bg}}
\affil[1]{\small \it Department of Theoretical Physics, Sofia University, Sofia 1164, Bulgaria}
\affil[2]{\it Institut f\"{u}r Physik, Universit\"{a}t Oldenburg, D-26111 Oldenburg, Germany}

\affil[3]{\it Theoretical Astrophysics, Eberhard-Karls University of T\"ubingen, T\"ubingen 72076, Germany}

\date{}

\maketitle

\begin{abstract}
We prove a uniqueness theorem for traversable wormhole solutions in the Einstein-Maxwell-dilaton gravity with a phantom scalar field and a possible phantom electromagnetic field. In a certain region of the parameter space, determined by the asymptotic values of the scalar field and the lapse function, the regular wormholes are completely specified by their mass, scalar charge and electric charge. The argument is based on the positive energy theorem applied on an appropriate conformally transformed Riemannian space.
\end{abstract}

\section{Introduction}

Wormholes arise naturally in general relativity and alternative theories of gravity as solutions  which form bridges between different universes, or different regions of the same universe \cite{MTV}. Physically most interesting are the so called traversable wormholes, which are free of conical or curvature singularities, and can be considered in context of future spacetime travel \cite{Teo}. Solutions with conical singularity localized in the equatorial plane (ring wormholes) also attract attention as models of elementary particles sourced by a cosmic string \cite{Clement}.

It is well known that in general relativity traversable wormholes cannot be supported by matter obeying the null energy condition \cite{V}. Its violation requires the existence of some exotic matter with negative energy density. Such scenarios are widely investigated also in cosmological context in view of dark energy models. One of the most simple models proposes  the introduction of  phantom matter fields, which possess kinetic terms coupled repulsively to gravity, such as phantom scalar or electromagnetic fields. Their presence agrees with current cosmological observations, and enables a variety of wormhole solutions \cite{EKK}. We should note, however, that regular wormholes can exist in certain alternative theories of gravity without introducing any exotic matter \cite{KK}.

In contrast to black holes, wormhole solutions, although available in many cases, are not classified systematically. A recent work proved that the Ellis-Bronnikov wormhole is the unique static traversable wormohole in Einstein-scalar field theory \cite{Yazadjiev:2017}. The goal of the present paper is to provide a uniqueness theorem for traversable wormholes in the static sector of the Einstein-Maxwell-dilaton theory with a phantom dilaton field and a possible phantom electromagnetic field. We consider the case of dilaton-Maxwell coupling constant equal to unity.

\section{Field equations and general definitions}

We consider Einstein-Maxwell-dilaton theory with a phantom dilaton field defined by  the following action

\begin{equation}\label{EMD_th}
{\cal S} = \frac{1}{16\pi}\int d x^4\sqrt{-g}\left[ \mathfrak{R} + 2\,{}^{\mathfrak{g}}\!\nabla_{\mu}\varphi\,{}^{\mathfrak{g}}\!\nabla^{\mu}\varphi -\varepsilon\,e^{-2\varphi} F_{\mu\nu}F^{\mu\nu}\right].
\end{equation}

\noindent
We further allow that the Maxwell-gravity coupling constant $\varepsilon$ can take either positive, or negative values $\varepsilon =\pm1$. Thus, we include in our analysis also phantom Maxwell fields. The action leads to the following field equations on the spacetime manifold $M^{(4)}$

\begin{align}\label{EMD_eq}
&\mathfrak{R}_{\mu\nu}=-2\,{}^{\mathfrak{g}}\!\nabla_{\mu}\varphi\,{}^{\mathfrak{g}}\!\nabla_{\nu}\varphi+ 2\varepsilon e^{-2\varphi}\left( F_{\mu\beta}F^{\ \beta}_{\nu} -\frac{\mathfrak{g}_{\mu\nu}}{4}F_{\beta\gamma}F^{\beta\gamma} \right), \\[2mm]
&{}^{\mathfrak{g}}\!\nabla_{[\beta}F_{\mu\nu]}=0,\nonumber \\[2mm]
&{}^{\mathfrak{g}}\!\nabla_{\beta}\left( e^{-2\varphi}F^{\beta\mu} \right)=0, \nonumber\\[2mm]
&{}^{\mathfrak{g}}\!\nabla_{\beta}\,{}^{\mathfrak{g}}\!\nabla^{\beta}\varphi=\frac{1}{2}\varepsilon e^{-2\varphi}F_{\mu\nu}F^{\mu\nu},\nonumber
\end{align}

\noindent where $\mathfrak{R}_{\mu\nu}$ is the spacetime Ricci tensor, $\varphi$ represents the dilaton field, and $F_{\mu\nu}$ is the Maxwell field.

We are interested in {\it strictly static spacetimes}, i.e. spacetimes possessing a Killing vector field $\xi$, which is everywhere timelike.
For such spacetimes there exists a smooth Riemannian manifold $(M^{(3)},g^{(3)})$ and a smooth lapse function $N:M^{(3)}\longrightarrow \mathbb{R}^+$ allowing the following decomposition
\begin{eqnarray}
M^4=\mathbb{R}\times M^{(3)},\quad~~~ \mathfrak{g}_{\mu\nu}dx^{\mu}dx^{\nu}=-N^2\mathrm{d}t^2+g^{(3)}_{ij}dx^idx^j.\label{statg}
\end{eqnarray}

\indent Staticity of the Maxwell and scalar fields is defined by means of the Lie derivative along the timelike Killing field $\xi=\frac{\partial}{\partial t}$,
\begin{align}
&\mathcal{L}_{\xi}F=0, \quad~~~\mathcal{L}_{\xi}\varphi=0.
\end{align}
We will further focus on the case of purely electric field,  which means that $\iota_{\xi}\star F=0$ is satisfied.

We consider static, asymptotically flat wormhole solutions to the field equations (\ref{EMD_eq}) with two asymptotically flat ends. In analogy to \cite{Yazadjiev:2017} we adopt the following formal definition:

\medskip
\noindent

{\it A solution to the field equations (\ref{EMD_eq}) is said to be  a static, asymptotically flat traversable  wormhole solution if the following conditions are satisfied:

\medskip
\noindent	
	
1. The spacetime is strictly static.

\medskip
\noindent

2. The Riemannian manifold $(M^{(3)},g^{(3)})$ is complete.

\medskip
\noindent

3.  For some compact set $K$, $M^{(3)}-K$ consists of two ends $End_{+}$ and $End_{-}$  such that each end is diffeomorphic to  $\mathbb{R}^{(3)} \backslash \bar{B}$ where $\bar{B}$ is the closed unit ball centered at the origin in $\mathbb{R}^{(3)}$, and with the following asymptotic behavior of the 3-metric, the lapse function, the scalar field, and the electromagnetic field  }

\begin{eqnarray}\label{asympt}
g^{(3)}_{ij} &=& N^{-2}_{\pm}\left(1 + \frac{2M_{\pm}}{r}\right)\delta_{ij} + {\cal O}(r^{-2}), \quad~~~ N= N_{\pm}\left(1 - \frac{M_{\pm}}{r}\right) + {\cal O}(r^{-2}), \nonumber\\
\varphi &=& \varphi_{\pm} - \frac{q_{\pm}}{r} + {\cal O}(r^{-2}), \quad~~~ F = -\, \left(\frac{Q_\pm}{r^2} + {\cal O}(r^{-3})\right) dt\wedge dr,
\end{eqnarray}

\noindent
{\it with respect to the standard radial coordinate $r$ of $\mathbb{R}^{(3)}$, where $\delta_{ij}$ is the standard flat metric on $\mathbb{R}^{(3)}$}.

\medskip

The conditions in the above wormhole definition are chosen in order to  reflect the expected properties of the wormhole geometry in a natural way. The first condition involves the requirement  that no horizons should be present, the second one ensures that the constant time slices are free from singularities, while the third one is the standard notion of asymptotically flat regions.

In our notations $N_{\pm}>0$, $M_{\pm}$, $\varphi_{\pm}\ne 0$,  $q_{\pm}\ne 0$, $\Phi_{\pm}\neq 0$, and $Q_{\pm}\ne 0$ are constants.  $M_{\pm}$ and $q_{\pm}$ represent the total (ADM) mass and the scalar charge of the corresponding end $End_{\pm}$. The parameters $Q_{\pm}$ are connected to the conserved electric charge associated with each end, however deviating from it by a multiplicative factor\footnote{The conserved electric charge is defined by eq. ($\ref{charge}$).}.
We choose $N_{-}\le N_{+}$.

{\it In the present paper we will also assume that the 3-dimensional manifold $M^{(3)}$ is simply connected}. We can introduce the electric field one-form
\begin{align}
E=-\iota_{\xi}F,
\end{align}
which satisfies $\mathrm{d}E=0$ as a consequence of the field equations and the staticity of the Maxwell field. Then, since
$M^{(3)}$ is simply connected,  there exists an electromagnetic potential $\Phi$ defined on $M^{(3)}$ such that $E=\mathrm{d}\Phi$. The Maxwell 2-form is given by
\begin{align}
F=-N^{-2}\xi\wedge\mathrm{d}\Phi,
\end{align}
and considering ($\ref{asympt}$) we obtain the asymptotic behavior of the potential

\begin{eqnarray}
\Phi = \Phi_\pm + \frac{Q_\pm}{r} + {\cal O}(r^{-2}).
\end{eqnarray}

Using the metric and the electromagnetic field decomposition, we can obtain the following dimensionally reduced static  EMD equations
\begin{align}\label{EDM_red}
&{}^g\!\Delta N=\varepsilon\, N^{-1}e^{-2\varphi}\,{}^g\!\nabla_i\Phi\,{}^g\!\nabla^i\Phi, \\[2mm]
&{}^g\!R_{ij}= -2\,{}^g\!\nabla_i\varphi\,{}^g\!\nabla_j\varphi+N^{-1}\,{}^g\!\nabla_i\,{}^g\!\nabla_jN \nonumber\\[2mm]
&\ \ \ \ \ \ \ \ +\varepsilon\,N^{-2}e^{-2\varphi}( g_{ij}\,{}^g\!\nabla_k\Phi\,{}^g\!\nabla^k \Phi -2\,{}^g\!\nabla_i\Phi \,{}^g\!\nabla_j\Phi ),\notag\\[2mm]
&{}^g\!\nabla_i(N^{-1}e^{-2\varphi}\,{}^g\!\nabla^i\Phi)=0,\nonumber\\[2mm]
&{}^g\!\nabla_i(N\,{}^g\!\nabla^i\varphi)=-\varepsilon\, N^{-1}e^{-2\varphi}\,{}^g\!\nabla_i\Phi\,{}^g\!\nabla^i\Phi,\nonumber
\end{align}
where ${}^g\!\nabla$ and ${}^g\!R_{ij}$ are the Levi-Civita connection and the Ricci tensor with respect to the 3-metric $g^{(3)}_{ij}$.

Using the maximum principle for elliptic partial differential equations and from
the asymptotic behaviour of $N$  it follows that the values of $N$ on $M^{(3)}$
satisfy

\begin{eqnarray}
N_{-}\le N\le N_{+}
\end{eqnarray}
as the equality is satisfied only in the case $Q_{\pm}=M_{\pm}=0$.

\section{Divergence identities and functional relations between the potentials  }

\indent We consider the conformally transformed metric $\gamma_{ij}$ on $M^{(3)}$ defined by

\begin{align}\label{gamma}
\gamma_{ij}=N^2g_{ij}.
\end{align}

\noindent
It is also convenient to introduce  further the potentials

\begin{eqnarray}\label{potentials}
U=\ln(N)+ \varphi, \quad~~~ \Psi= \ln(N)-\varphi,
\end{eqnarray}
allowing to simplify the reduced field equations ($\ref{EDM_red}$) to the form

\begin{eqnarray}\label{FE_PS}
&{}^{\gamma}\!R_{ij}= D_iUD_j\Psi +  D_i\Psi D_jU-2{ \varepsilon}\, e^{-2U}D_i{\Phi}D_j{\Phi}, \\[2mm]
&D_iD^iU= 0,  \nonumber \\[2mm]
&D_iD^i\Psi= 2\varepsilon\,e^{-2U}D_i{\Phi}D^i{\Phi}, \nonumber \\[2mm]
&D_i(e^{-2U}D^i{\Phi})=0. \nonumber
\end{eqnarray}
where $D_i$ denotes the covariant derivative with respect to the metric $\gamma_{ij}$.

For a given solution $(\gamma_{ij},U,\Psi, \Phi)$  the transformation ($\gamma_{ij}\to \gamma_{ij}$, $U\to U+ C_1$, $\Psi\to \Psi+ C_2$, $\Phi\to e^{C_1}(\Phi +C_{3})$)  where $C_1$, $C_2$
and $C_{3}$ are constants, generates new solutions. These new solutions can however be regarded as trivial. In order to get rid of them we
shall impose the following constraints on the potentials $U$, $\Psi$ and $\Phi$:

\begin{eqnarray}\label{ACP}
U_{+}=-U_{-}, \; \Psi_{+}=- \Psi_{-}, \;\Phi_{+}=-\Phi_{-} .
\end{eqnarray}

The field equations ($\ref{FE_PS}$) can be interpreted as describing a 3-dimensional gravity coupled to a non-linear $\sigma$-model parameterized by the scalar fields $\phi^A=(U,\Psi,\Phi)$ with a target space  metric
\begin{align}\label{TSMETRIC}
G_{AB}d\phi^A d\phi^B=\frac{1}{2}(\mathrm{d}U\mathrm{d}\Psi+\mathrm{d}\Psi\mathrm{d}U) -  2\varepsilon e^{-2U}\mathrm{d}{\Phi}^2.
\end{align}

\noindent The Killing vectors for this metric are
\begin{align}
&K^{(1)}=2\varepsilon\,{\Phi}\frac{\partial}{\partial \Psi}+ \frac{1}{2}e^{2U}\frac{\partial}{\partial {\Phi}},\label{kill1}\\
&K^{(2)}=\frac{\partial}{\partial U}+{\Phi}\frac{\partial}{\partial {\Phi}},\nonumber \\
&K^{(3)}=-\varepsilon\frac{\partial}{\partial {\Phi}},\nonumber \\
&K^{(4)}=\frac{\partial}{\partial \Psi},\nonumber
\end{align}
and the  corresponding Killing one-forms are given by
\begin{align}
&K^{(1)}_{A}d\phi^A =2{\Phi}dU - d{\Phi},\label{killf1} \\
&K^{(2)}_{A}d\phi^A= - d\Psi +  2\varepsilon\,e^{-2U}{\Phi}d{\Phi},\nonumber \\
&K^{(3)}_{A}d\phi^A= e^{-2U}d{\Phi}, \nonumber\\
&K^{(4)}_{A}d\phi^A=dU. \nonumber
\end{align}

\indent Using the fact that $K^{(a)}_A d\phi^A $ are Killing one-forms for the metric $G_{AB}$ and taking into account the  equations for $\phi^A$ one can show that the following divergence identities are satisfied
\begin{align}
D_i(K^{(a)}_AD^i\phi^A)=0.
\end{align}

\noindent We integrate these equations over $M^{(3)}$ and consider the behaviour of the potentials $\phi^A$ at the two asymptotically flat ends  $End_{+}$ and $End_{-}$  given by ($\ref{asympt}$). Thus, the following relations are obtained
\begin{eqnarray}\label{ID}
&&  M_{+} + q_{+} = - M_{-} - q_{-}, \\[2mm]
&&e^{-2U_{+}}\,Q_{+} = -e^{-2U_{-}}\,Q_{-}, \nonumber  \\[2mm]
&&M_{+} - q_{+} = - (M_{-} -  q_{-}) - 2\varepsilon\tilde{Q}_+\Delta{\Phi}, \nonumber\\[2mm]
&&e^{2U_{+}}=e^{2U_{-}}-2(M_{+}+ q_{+})\frac{\Delta{\Phi}}{{\tilde Q_+}}, \label{PR}
\end{eqnarray}

\noindent
where we have introduced the notations

\begin{eqnarray}\label{charge}
\tilde{Q}_{\pm} &=& e^{-2U_{_{\pm}}}\,Q_{_{\pm}},  \\[2mm]
\Delta{\Phi} &=& {\Phi}_{+}-{\Phi}_{-}.
\end{eqnarray}

The quantities $\tilde Q_{\pm} = e^{-2U_{\pm}}\,Q_{\pm}$ represent the conserved electric charges associated with the two asymptotic ends.
The identities ($\ref{ID}$)-($\ref{PR}$) lead to Smarr-like relations connecting the conserved charges for the two asymptotic ends

\begin{eqnarray}
&&M_{+} + M_{-} + \varepsilon\tilde{Q}_+\Delta{\Phi} =0, \nonumber \\[2mm]
&&q_{+} +  q_{-} - \varepsilon\tilde{Q}_+\Delta{\Phi}=0.
\end{eqnarray}
Thus, we can retain as independent parameters, which characterize the wormhole solutions, only the asymptotic charges associated with $End_{+}$.

We can expand the last relation ($\ref{PR}$) between the asymptotic properties to a functional relation valid on the whole 3-dimensional manifold $M^{(3)}$. We consider the potential

\begin{align}
\chi_1 = e^{2U_{+}}- e^{2U}-2(M_{+}+ q_{+})\frac{\tilde{\Phi}}{{\tilde Q}_{+}},
\end{align}
where we denote  $\tilde{\Phi} = {\Phi} - {\Phi}_{+}$, which is normalized appropriately to vanish at both asymptotic ends  $End_\pm$. Then, we can construct a related 1-form $\omega = - e^{-2U}d\chi_1$, given explicitly by the expression

\begin{eqnarray}
\omega=2dU+2e^{-2U} \frac{(M_{+} + q_{+})}{\tilde Q_{+}} d{\Phi},
\end{eqnarray}

\noindent
which obeys the relation

\begin{eqnarray}
e^{2U}\omega_i\omega^i = - D_i\chi_1\,\omega^i = -D_i(\chi_1\omega^i),
\end{eqnarray}

\noindent
as a result of the field equations. Integrating this relation over $M^{(3)}$ we obtain

\begin{eqnarray}
\int_{M^{(3)}} e^{2U}\omega_i\omega^i\mathrm{d}\mu= - \int_{M^{(3)}}D_i(\chi_1\omega^i)\mathrm{d}\mu = -\int_{S^\infty_{+}\bigcup S^{\infty}_{-}}\chi_1\omega^i dS_i = 0,
\end{eqnarray}

\noindent
where we have used the asymptotic behaviour of the potential $\chi_1$. Then, it is satisfied that $\omega_i=0$  on $M^{(3)}$, and consequently the potential $\chi_1$ vanishes identically. Hence, we obtain the following relation between the potentials $U$ and $\Phi$

\begin{align}\label{chi}
 e^{2U_{+}}- e^{2U}-2\tilde M_+\frac{\tilde{\Phi}}{{\tilde Q}_{+}}  =0,
\end{align}
where we introduce the notation $\tilde M_+ = M_{+}+ q_{+}$.

We can further consider the potential $\eta$, defined by
\begin{eqnarray}
d\eta =  d\Psi - 2\varepsilon e^{-2U}\Phi d\Phi,
\end{eqnarray}
and construct the current $J_i= U D_i \eta-\eta D_i U$. The current $J_i$ is conserved as a consequence of the field equations, i.e. $D_i J^i=0$ on $M^{(3)}$. Integrating this equation over $M^{(3)}$, and using the asymptotic identities ($\ref{ID}$), we get
\begin{eqnarray}\label{Psi}
\left[(M_{+} - q_{+})+ 2\varepsilon\Phi_+\tilde Q_+\right]\Delta U-\tilde M_{+}\Delta\eta=0,
\end{eqnarray}

\noindent
where the quantities $\Delta U = U_{+} - U_{-}=2U_{+}$ and $\Delta\eta = \eta_{+} - \eta_{-}$ are defined by means of the values of the potentials $U$ and $\eta$ at the two asymptotic ends $End_\pm$. The identity can be extended to a relation valid on the whole $M^{(3)}$. We introduce the potential

\begin{eqnarray}
\chi_2 = \left[(M_{+} - q_{+})+ 2\varepsilon\Phi_+\tilde Q_+\right]\tilde U-\tilde M_{+}\tilde\eta,
\end{eqnarray}

\noindent
where we denote $\tilde U = U - U_{+}$, $\tilde\eta = \eta - \eta_{+}$, which by construction vanishes on both asymptotic ends. We can further consider the 1-form $\varpi = d\chi_2$, satisfying
\begin{eqnarray}
\varpi_i\varpi^i = D_i\chi_2\varpi^i = D_i(\chi_2\varpi^i),
\end{eqnarray}

\noindent
as a result of the field equations. Integrating this equation over $M^{(3)}$ and using the asymptotic behavior of the potential $\chi_2$, we get that $\varpi_i = 0$ on $M^{(3)}$, and consequently $\chi_2$ also vanishes identically. Hence, we obtain the relation

\begin{eqnarray}
 \left[(M_{+} - q_{+})+ 2\varepsilon\Phi_+\tilde Q_+\right]\tilde U =\tilde M_{+}\tilde\eta
\end{eqnarray}

\noindent
which leads to

\begin{align}\label{Psi_zeta}
\Psi = -\frac{q^2_{+} -M^2_{+} + \varepsilon {\tilde Q}^2_{+}e^{2U_{+}} }{{\tilde M}^2_{+}} U
+ \frac{\varepsilon {\tilde Q}^2_{+}}{2{\tilde M}^2_{+}} \left( e^{2U} - \frac{1}{2}e^{2U_{+}} - \frac{1}{2}e^{-2U_{+}} \right).
\end{align}

Using the relations $\Phi(U)$ and $\Psi(U)$ we can express the field equations ($\ref{FE_PS}$) by means of a single potential $U$ in the form

\begin{align}\label{FE_U}
&{}^{\gamma}\!R_{ij}= - \frac{2}{{\tilde M}^2_{+}}\left(q^2_{+} -M^{2}_{+} + \varepsilon {\tilde Q}^2_{+} e^{2U_{+}} \right)D_i U D_j U,\\
&D_iD^i U=0. \nonumber
\end{align}

\indent Using the field equations we can derive an inequality restricting the asymptotic values of potential $U$ and ${\tilde M}_{+}$ at the asymptotic ends $End_{\pm}$
\begin{align}\label{zeta_Q}
\int_{M^3}D_iU D^i U\mathrm{d}\mu =& \int_{M^3}D_i(U D^i U) \mathrm{d} \mu \nonumber \\
=&\int_{S^{\infty}_+}U D_i U\mathrm{d}S^i+\int_{S^{\infty}_-}U D_i U \mathrm{d}S^i \nonumber \\
=&4\pi\left( U_+ - U_- \right)\tilde{M}_{+}= 8\pi U_{+}\tilde{M}_{+} >0.
\end{align}

\section{Uniqueness theorem}

We will prove a uniqueness theorem for the static asymptotically flat wormhole solutions in the Einstein-Maxwell-dilaton theory given by ($\ref{EMD_th}$), with a phantom scalar field and a possible phantom electromagnetic field. The main steps of the proof are based on the
ideas developed in \cite{Yazadjiev:2017}. The argument is valid only for a certain range of the solution parameters satisfying a constraint given below.

The following statement can be formulated:

\medskip
\noindent

{\bf Theorem:} {\it The asymptotic charges $M_{+}$, $q_{+}$, ${\tilde Q}_{+}$ and the asymptotic value $U_{+}$ of the potential $U$ for static, asymptotically flat traversable wormhole solutions to  the  Einstein-Maxwell-dilaton equations ($\ref{EMD_eq}$) with a  phantom scalar field and a possible phantom electromagnetic fields satisfy  the inequality

\begin{eqnarray}
 q_{+}^2 - M_{+}^2  + {\varepsilon}\tilde Q_{+}^2e^{2U_+}  >0.
\end{eqnarray}

Moreover, for fixed value of the parameter $\varepsilon = \pm 1$ there can be only one static, asymptotically flat traversable wormhole spacetime $(M^{(4)}, g^{(4)},\varphi, \Phi)$ with asymptotic values of the potentials satisfying (\ref{ACP}), with given mass $M_+$,  scalar charge $q_+$, electric charge $\tilde Q_+$, and asymptotic value of the potential $U_+$  satisfying the inequality

\begin{eqnarray}\label{constr2}
0< \sqrt{\frac{q^2_{+} -M^{2}_{+} + \varepsilon {\tilde Q}^2_{+} e^{2U_{+}}}{{\tilde M}^2_{+}} }\, U_+ \leq \frac{\pi}{2} .
\end{eqnarray}

It is isometric to the spherically symmetric solution constructed below. }

{\bf Proof:} We consider the three-dimensional Riemannian manifold $(M^{(3)},\gamma_{ij})$ with metric $\gamma_{ij}$ given by ($\ref{gamma}$). It  is a complete asymptotically flat manifold with two ends possessing vanishing mass, ${}^{\gamma}M_{\pm}=0$. The completeness of the metric $\gamma_{ij}$ is ensured by its definition, combined with the fact that the metric $g^{(3)}_{ij}$ is complete and the lapse function $N$ is bounded on $M^{(3)}$.

 The asymptotic behavior of $\gamma_{ij}$ can be obtained from the asymptotic behavior of $g^{(3)}_{ij}$ and the lapse function $N$ as

\begin{eqnarray}
\gamma_{ij}= \delta_{ij} + O(r^{-2}).
\end{eqnarray}

Let us assume  that the following inequality between the conserved charges and the asymptotic values of the potentials is satisfied

\begin{eqnarray}\label{ineq1}
q_{+}^2 - M_{+}^2  + {\varepsilon}\tilde Q_{+}^2 e^{2U_+} \leq 0.
\end{eqnarray}

\noindent
Then we have a complete asymptotically flat Riemannian manifold  $(M^{(3)}, \gamma_{ij})$,  which  possesses a non-negative scalar curvature as can be seen from   ($\ref{FE_U}$), and zero total mass for each of its ends. From the rigidity of the positive energy theorem \cite{SYB}  it follows that   $(M^{(3)},\gamma_{ij})$ is isometric to $(R^{(3)}, \delta_{ij})$. Thus, our assumption that  ($\ref{ineq1}$) is satisfied leads to  a contradiction. Therefore, we conclude that for wormhole solutions  the mass, the electric charge and the scalar charge have to satisfy
the inequality

\begin{eqnarray}\label{constr1}
q_{+}^2 - M_{+}^2  + {\varepsilon}\tilde Q_{+}^2 e^{2U_+}>0.
\end{eqnarray}

Under this condition we can introduce a new scalar field $\lambda$, defined by

\begin{eqnarray}\label{lambda_def}
\lambda=  \sqrt{\frac{q^2_{+} -M^{2}_{+} + \varepsilon {\tilde Q}^2_{+} e^{2U_{+}}}{{\tilde M}^2_{+}} }\, U,
\end{eqnarray}
 with asymptotic values constrained in the range
\begin{eqnarray}\label{lambda_asympt}
0<\lambda_+ = -\lambda_- \leq \frac{\pi}{2},
\end{eqnarray}
as follows from (\ref{constr2}).

\noindent

In terms of the new potential $\lambda$ the field equations $\label{FE_zeta}$ take the form

\begin{eqnarray}\label{EFFEL}
&&R(h)_{ij}= -2D_{i}\lambda D_{j}\lambda, \\
&&D_{i}D^{i}\lambda = 0. \nonumber
\end{eqnarray}

\noindent
In this way we reduce the problem to the one solved in \cite{Yazadjiev:2017}. Under the asymptotic condition ($\ref{lambda_asympt}$) we can perform the  conformal transformation

\begin{eqnarray}
h_{ij}= \Omega^2 \gamma_{ij},
\end{eqnarray}
where the conformal factor is given by
\begin{eqnarray}
	\Omega^2 = \frac{\sin^4(\frac{\lambda + \lambda_{+} }{2})}{\sin^4(\lambda_{+})} .
\end{eqnarray}
Using the field equations ($\ref{EFFEL}$) we can show that the scalar curvature of the conformably transformed metric $h_{ij}$ vanishes. The conformally transformed metric can be expanded as

\begin{eqnarray}
{\gamma}_{ij}=\Omega^2 h_{ij}= \frac{\left(q_{+}^2 -M_{+}^2+\varepsilon_2e^{2U_+}\tilde Q_{+}^2\right)^2}{16 \sin^4(\lambda_{+}) r^4} \delta_{ij} + O(1/r^6)
\end{eqnarray}
\noindent
near $End_{-}$ in the standard asymptotic coordinate $r\to \infty$. We can perform a coordinate transformation $y^i= x^i/r^2$, and introduce a new radial coordinate $R$ such that $R^2=\delta_{ij}y^i y^j$. Then, the asymptotic expansion of the metric near $End_{-}$ is performed when $R\to 0$, and we obtain

\begin{eqnarray}
{h}(\frac{\partial}{\partial y^i},\frac{\partial}{\partial y^j})= \frac{\left(q_{+}^2 -M_{+}^2+\varepsilon_2e^{2U_+}\tilde Q_{+}^2\right)^2}{16 \sin^4(\lambda_{+})} \delta_{ij} + O(R^2).
\end{eqnarray}

Consequently, we can add a point $\infty$ at $R=0$, and construct a sufficiently regular manifold ${\tilde M^{(3)}}=M^{(3)} \cup \infty$.
By construction the Riemannian manifold ${\tilde M^{(3)}}=M^{(3)} \cup \infty$ is geodesically complete, scalar flat manifold  with one asymptotically flat end $End_{+}$. According to the positive energy theorem \cite{SYB} its total mass ${\tilde M}^{(h)}$ with respect to the metric $h_{ij}$ should be non-negative, i.e.  ${\tilde M}^{(h)}\ge 0$ should be satisfied. We can obtain the mass ${\tilde M}^{(h)}$ from the asymptotic behaviour of $h_{ij}$, which is given by

\begin{eqnarray}
h_{ij}= \left(1- 2\cot(\lambda_{+})\sqrt{q_{+}^2 -M_{+}^2+\varepsilon_2e^{2U_+}\tilde Q_{+}^2}\,\frac{1}{r}\right)\delta_{ij} + O(r^{-2}),
\end{eqnarray}
when $r\to \infty$. Hence we have ${\tilde M}^{(h)}= -2\cot(\lambda_{+})\sqrt{q_{+}^2 -M_{+}^2+\varepsilon_2e^{2U_+}\tilde Q_{+}^2}$. Taking into account the inequality (\ref{zeta_Q}) it follows that  ${\tilde M}^{(h)}\le 0$, as the equality is saturated only for $\lambda_{+}=\frac{\pi}{2}$. Therefore, we obtain that  ${\tilde M}^{(h)}=0$ and $\lambda_{+}=\frac{\pi}{2}$. Let us note that
$\lambda_{+}=\frac{\pi}{2}$  through the relation ($\ref{lambda_def}$) gives an algebraic relation which determines $U_+$ as a function of
$M_{+}$, $q_{+}$ and ${\tilde Q}_{+}$.

In summary, we constructed a  geodesically complete, scalar flat Riemannian manifold $({\tilde M^{(3)}}, h_{ij})$ with one asymptotically flat end and vanishing total mass. Then, by  the positive energy  theorem in the rigid case \cite{SYB} it follows that $({\tilde M^{(3)}},h_{ij})$ is isometric to
$(\mathbb{R}^3, \delta_{ij})$. Consequently,  the metrics $\gamma_{ij}$ and $g^{(3)}_{ij}$  are conformally flat and $M^{(3)}$ is diffeomorphic to $\mathbb{R}^3/\{0\} $.

The wormhole solutions satisfying the theorem assumptions can be constructed by straightforward integration of the field equations (\ref{EFFEL}) in spherical coordinates. More precisely we have to integrate   (\ref{EFFEL})  for the metric

\begin{eqnarray}
\gamma_{ij}dx^idx^j= \sin^{-4}(\frac{\lambda }{2} + \frac{\pi }{4} )(dR^2 + R^2 d\theta^2 + R^2\sin^2\theta d\phi^2),
\end{eqnarray}
with the  asymptotic conditions $-\frac{\pi}{2}\leq\lambda\leq \frac{\pi}{2}$. As a result,  we obtain the solution

\begin{eqnarray}
&&\gamma_{ij}dx^idx^j= \left(1+ \frac{q_{+}^2 -M_{+}^2+\varepsilon_2e^{2U_+}\tilde Q_{+}^2}{4R^2}\right)^2 \left(dR^2 + R^2 d\theta^2 + R^2\sin^2\theta d\phi^2\right), \nonumber \\[2mm]
&&\lambda= 2\arctan(\frac{2R}{\sqrt{q_{+}^2 -M_{+}^2+\varepsilon_2e^{2U_+}\tilde Q_{+}^2}}) -\frac{\pi}{2},
\end{eqnarray}
where $R\in(0,+\infty)$. We can further introduce another radial coordinate $x$ taking the symmetric range $x\in(-\infty,\infty)$ by  the coordinate transformation $x= R- \frac{q_{+}^2 -M_{+}^2+\varepsilon_2e^{2U_+}\tilde Q_{+}^2}{4R}$, and write the metric in the form

\begin{eqnarray}
&&\gamma_{ij}dx^idx^j= dx^2 + (x^2 + q_{+}^2 -M_{+}^2+\varepsilon_2e^{2U_+}\tilde Q_{+}^2)(d\theta^2 + \sin^2\theta d\phi^2), \nonumber \\[2mm]
&&\lambda=\arctan(\frac{x}{\sqrt{q_{+}^2 -M_{+}^2+\varepsilon_2e^{2U_+}\tilde Q_{+}^2}}).
\end{eqnarray}	

By construction we have obtained the unique form of the potential $\lambda$ corresponding to a regular wormhole solution, and satisfying the prescribed asymptotic behavior. It determines the potential $U$ by the relation ($\ref{lambda_def}$), from which we can generate explicitly the wormhole solutions as described in the following section.

\subsection{Construction of the EMD  phantom wormhole spacetimes}

The explicit form of the solutions, which satisfy the presented uniqueness theorem can be obtained by using the relations $\Phi(U)$ and $\Psi(U)$ given by ($\ref{chi}$) and ($\ref{Psi_zeta}$), as well as the definition ($\ref{lambda_def}$). Hence, we obtain the electromagnetic field as a function of the potential $\lambda$

\begin{eqnarray}\label{Phi1}
\Phi &=& -\frac{\tilde Q_+}{2\tilde M_+}\left[ e^{\frac{2\tilde M_+\lambda}{\,C_\lambda}}-\cosh{\frac{\pi\tilde M_+}{\,C_\lambda}}\right], \\[2mm]
C_\lambda &=& \sqrt{q_{+}^2 -M_{+}^2+\varepsilon_2e^{2U_+}\tilde Q_{+}^2}\,. \nonumber
\end{eqnarray}

The metric function $N$ and the dilaton field $\varphi$ are further extracted from the potentials $U$ and $\Psi$ using their definition ($\ref{potentials}$)

\begin{eqnarray}\label{sol1}
N^2&=& \exp\left[\left(2M_+ -  \frac{\varepsilon\,\tilde Q^2_+}{\tilde M_+}e^{2U_+}\right)\frac{\lambda }{C_\lambda} + \frac{\varepsilon \,\tilde Q^2_+}{2\tilde M^2_+}\left(e^{\frac{2\tilde M_+\lambda}{\, C_\lambda}} - \cosh {\frac{\pi\tilde M_+}{\,C_\lambda}}\right)\right], \nonumber \\[2mm]
\varphi &=& \left(q_+ +  \frac{\varepsilon\,\tilde Q^2_+}{2\tilde M_+} e^{2U_+}\right)\frac{\lambda}{C_\lambda} - \frac{\varepsilon \,\tilde Q^2_+}{4\tilde M^2_+}\left(e^{\frac{2\tilde M_+\lambda}{\, C_\lambda}} - \cosh {\frac{\pi\tilde M_+}{\,C_\lambda}}\right).
\end{eqnarray}

In the case of positive Maxwell-gravity coupling constant $\varepsilon=1$ the constructed solution was previously obtained by a different procedure in \cite{Goulart:2017}. In this work the solution is parameterized by the integration constants $b_1$, $b_2$, $c_1$, $c_2$ and $Q$,  which are related to our parameters $M_+$,  $q_+$, $\tilde Q_+$ and $U_+$  as
\begin{eqnarray}
c_1 &=& \tilde M_+, \quad~~~ c_2 = 0, \quad~~~ Q = -\tilde Q_+,  \nonumber  \\[2mm]
b_1 &=& q_+ + \frac{\tilde Q^2_+}{2\tilde M_+}e^{2U_+}, \quad~~~ b_2 =  \frac{\varepsilon \,\tilde Q^2_+}{4\tilde M^2_+}\cosh {\frac{\pi\tilde M_+}{\,C_\lambda}},
\end{eqnarray}
while the notation $l$ corresponds to our expression $C_\lambda$. The proof of the uniqueness theorem imposes a restriction on the asymptotic value of the potential $\lambda$, i.e. $\lambda_+ = -\lambda_- = \frac{\pi}{2}$, which through the relation ($\ref{lambda_def}$) leads to a restriction on the asymptotic value $U_+$. It should satisfy the algebraic equation

\begin{eqnarray}
 U_+ = \frac{\pi}{2}\sqrt{\frac{{\tilde M}^2_{+}}{q^2_{+} -M^{2}_{+} + \varepsilon {\tilde Q}^2_{+} e^{2U_{+}}}},
\end{eqnarray}
which can be always ensured,  since $M_+$,  $q_+$, $\tilde Q_+$ can take arbitrary values. Thus, we obtain that the solution is parameterized by three independent parameters, given by the conserved charges $M_+$,  $q_+$, $\tilde Q_+$ associated with the asymptotic $End_+$. By our theorem, in the parameter range specified by ($\ref{constr1}$), they determine the solution uniquely, and any regular traversable wormhole should coincide with the constructed solution ($\ref{Phi1}$)-($\ref{sol1}$).

\section*{Acknowledgements}
The support by the Bulgarian NSF Grant DFNI T02/6, Sofia University Research Fund under Grants 80.10-30/2017 and 3258/2017, COST Action MP1304, COST Action CA15117 and COST Action CA16104  is gratefully acknowledged.

\end{document}